# Towards a Practical Ethics of Generative AI in Creative Production Processes

Geert Hofman[1]

## Abstract

The increasing use of artificial intelligence (AI) in various domains, including design and creative processes, raises important ethical questions. While much has been written about AI ethics from the perspective of technology developers, less attention has been given to the practical ethical considerations for technology consumers, particularly in the context of design. This paper proposes a framework for navigating the ethical challenges of using AI in creative production processes, such as the double diamond design model. Drawing on six major ethical theories - virtue ethics, deontology, utilitarianism, contract theory, care ethics, and existentialism - we develop a "compass" for zooming in and out on the ethical aspects of AI in the design process. The framework emphasizes the importance of responsibility, anticipation, and reflection at each stage of the AI lifecycle as well as at each step of the creative process. We argue that by adopting a playful and exploratory approach to AI, while remaining grounded in fundamental ethical principles, designers can harness the potential of these technologies in a responsible and value-aligned way without over-burdening or compromising the creative process to much.

## Keywords

## Table of contents

---

[1] Department Business and Media, Network Economics, University College of West-Flanders, Kortrijk, Belgium

# Introduction

As artificial intelligence (AI) technologies become more powerful and pervasive, they are increasingly being used in a wide range of domains, including design and creative processes. The creative sector has always been at the forefront of adopting new technologies and methodologies, but with the advent of generative AI, capable of autonomously generating new images, texts, and concepts, there is talk of a potential paradigm shift in how creative professionals will do their work. Recent developments raise concerns about the possibility of machines replacing human creativity, and the role of the designer becoming subordinate to the role of the machine.

AI offers opportunities for creative professions, but there are also many challenges that need to be investigated to implement its application in a balanced and responsible manner. Finding a balance between human and machine creativity, ensuring human involvement during the design process, and addressing ethical, legal, and ecological implications are crucial issues to analyze before fully integrating the applications into daily business operations.

Much of the existing literature on AI ethics has focused on the responsibilities of technology developers, with less attention given to the practical ethical considerations for technology consumers, like professional designers in diverse creative industries, who mainly use AI in prepackaged configurable tools[2]. This paper seeks to address this gap by proposing a framework for navigating the governance and ethical challenges of using AI in creative production processes, such as the double diamond design model[3].

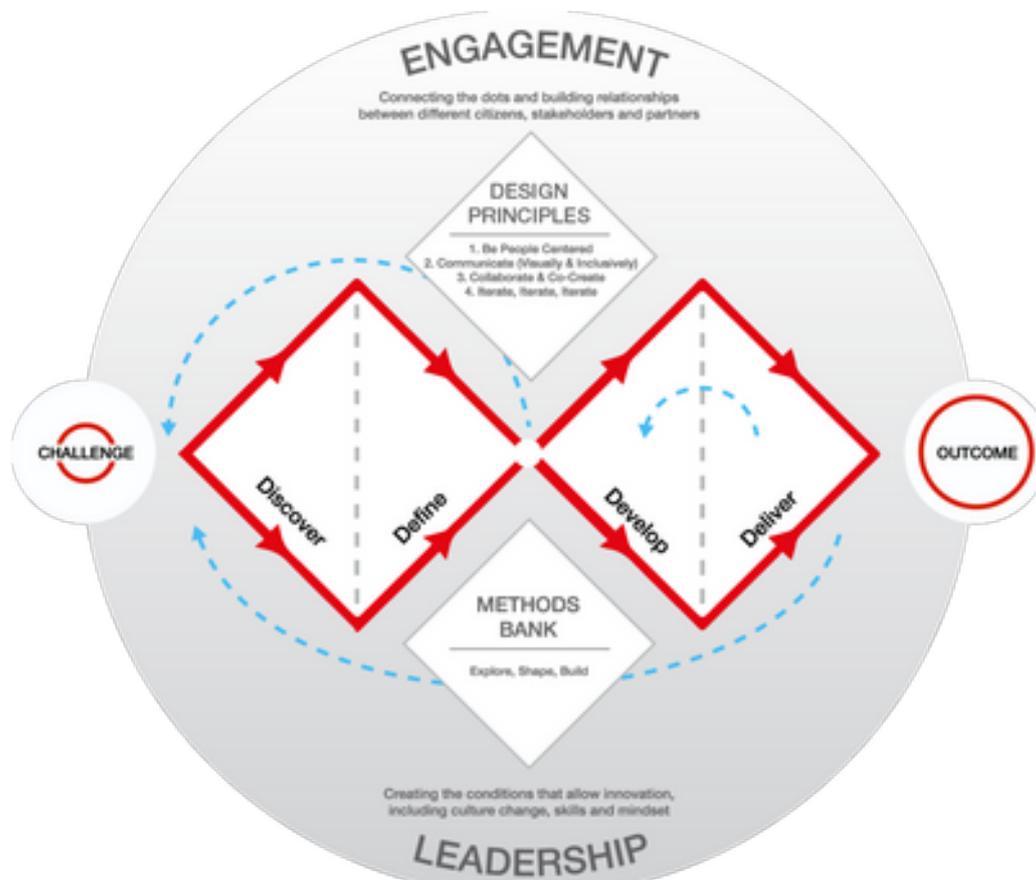

This work by the Design Council is licensed under a CC BY 4.0 license.

---

2  Ilina Georgieva et al., 'From AI Ethics Principles to Data Science Practice: A Reflection and a Gap Analysis Based on Recent Frameworks and Practical Experience', *AI and Ethics* 2, no. 4 (1 November 2022): 697–711, https://doi.org/10.1007/s43681-021-00127-3.
3  'Framework for Innovation - Design Council', accessed 7 June 2024, https://www.designcouncil.org.uk/our-resources/framework-for-innovation/.

The first part of this article tries to sketch the problem at hand. By focusing on the role of an informed team and possibly also on the need of a coordinating figure taking up the direction of the teams' creative effort, a major prerequisite for the responsible use of AI informed creative processes, is outlined. In the same sense as normally one supposes that designers know their jobs by using the right tools for the right purpose in the design process, one may suppose that this is also the case for the governance and ethical aspects in it's relation to the use of AI. The education, formation and governance of the team is the first crucial part of the ability of a creative team to navigate AI ingested design processes.

The second part focuses on the different elements needed to supply an informed creative team with the necessary tools to be able to use ethics as an added value in the creative process. There are three elements important in this context. Firstly to be able to handle the upstream ethical burden from AI tools, labeling is very important. Secondly, using existing ethical schools and theories as lenses to zoom in and out of ethical and governance aspects related to the task, can help getting a handle on things. Lastly, no solutions are definitive; not the creative solution itself nor the ethical aspects of it. It is therefore crucial that an attitude of playfulness and openness to future developments is present.

# Ethics for AI in design

## The team

In an era where generative AI is rapidly reshaping the creative industry – encompassing creative agencies, web agencies, product designers, marketeers, and communication professionals – the role of small and medium enterprises (SMEs) is becoming increasingly pivotal. Research into AI governance and ethics can help creating a framework or compass for balancing innovation with responsibility.

Generative AI offers teams in the creative sector unprecedented opportunities, from personalized design solutions to AI-driven marketing strategies. However, the integration of AI also brings forth significant challenges. The potential for AI-generated content to inadvertently embody biases or inaccuracies necessitates a robust governance structure. Integrating ethical considerations into AI governance is essential for ensuring fairness, transparency, and accountability.

Moreover, the creative sector often deals with intellectual property and originality – areas where AI's impact is still being understood. This underscores the need for research that defines clear ethical guidelines and governance models to navigate these uncharted waters.

Another critical aspect is data privacy and security, especially for SMEs handling sensitive client information. As AI systems process vast amounts of data, SMEs must ensure compliance with evolving data protection regulations such as the European data (protection) act and AI act. Integration of AI governance guidelines with existing organizational structures therefore becomes crucial.

Furthermore, as the creative sector is known for its innovation and flexibility, SMEs must also consider the cultural impact of AI integration. Research in AI governance should thus include strategies for maintaining a human-centric approach, ensuring that AI tools augment rather than replace human creativity and judgment.

As there are risks associated with the over-reliance on AI, including the potential loss of critical thinking and creativity skills, which are core to the creative sector, ethical AI governance should promote a balanced approach, leveraging AI's capabilities while nurturing human talent and creativity.

In this article opportunities for teams in the creative sector are sketched, showing them that AI governance and ethics, not only is becoming a regulatory necessity but also a strategic imperative. It ensures that while they leverage AI for innovation and competitive advantage, they also uphold ethical standards and contribute positively to the societal and economic fabric. A well-defined AI governance framework is instrumental in navigating the complex landscape of AI ethics and ensuring responsible, fair, and transparent AI use in the creative sector. It should be part of every creative teams' toolbox.

In large teams it won't always be possible to make every team member sufficiently ethically informed and responsible. In these cases it is advisable to give one or more team members the extra role of ethical conductor (in the sense of an orchestral conductor). This conductor (or conductor sub-team) can guide the rest of the team on a responsible path using the necessary tools as described in the next section.

## Use, think, try

### The importance of labeling (use)

The AI lifecycle encompasses several critical stages: creation, production, implementation, instruction, consumption, and policymaking. While other divisions within an AI application's lifecycle are certainly possible, these stages have the most significant impact when considering questions of responsibility[4]. At each stage, there are different stakeholders with potentially conflicting interests and ethical considerations. A key challenge is anticipating and mitigating negative consequences while promoting positive outcomes. This needs to be done taking the life cycle stage where the design process is in, into account.

When talking about creative agencies, product designers and marketeers or communication designers, mostly SMEs, the target group of our research, this is primarily the consumption stage of AI, that is, using and configuring AI tools. In such a case there are a lot of upstream governance and ethical considerations been thrown into the lap of the creative designer. Clearly the creative designer cannot be held responsible for all aspects of the AI tool. To be able to handle this upstream legacy, the use of labels for certain important ethical criteria can help relieve the professional end user of part of the ethical burden. These would need to be agreed upon by as many stakeholders involved in AI as possible. There should be representatives from industry, academia, government and non-governmental organisations validating these labels with an as wide as possible geographic distribution. In the food industry the Nutriscore and Eco-score labels are well known examples[5].

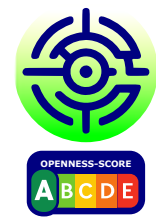

*Starcoder*

The primary labels that could be used in the context of AI and that are sufficiently aligned with attainable SMART[6] goals, are transparency and sustainability. It should be relatively easy to agree on relevant criteria for a decent and widely supported label. In an upcoming experiment with some of our user group we will deploy a proper label for transparency. To score that label we will be using the Foundation Model Transparency Index May 24 edition[7]. Each foundational model used in that index gets a percentage score. This is easily translatable to a label with 5 qualifications (A/B/C/D/E, with A being the best qualification). Next to this paragraph you can find an example of how this might look like.

The table below gives an overview of all tested model in the index.

---

4  Mark Haakman et al., 'AI Lifecycle Models Need to Be Revised: An Exploratory Study in Fintech', *Empirical Software Engineering* 26, no. 5 (September 2021): 95, https://doi.org/10.1007/s10664-021-09993-1.

5  Nutriscore is an initiative of the French department of health adopted by 7 states to indicate the nutritional value of food products. See Santé publique - France, 'Transnational governance of Nutri-Score: the 7 engaged countries adopt an improved algorithm for food', 29 July 2022, https://www.santepubliquefrance.fr/en/transnational-governance-of-nutri-score-the-7-engaged-countries-adopt-an-improved-algorithm-for-food for more info.
The Ecoscore tries to do something similar but is an initiative of a private organisation, namely Open Food Facts, 'Eco-Score: The Environmental Impact of Food Products', 2023, https://world.openfoodfacts.org/eco-score-the-environmental-impact-of-food-products.

6  A label is comparable with a goal setting tool. The aim of every scored label is to reach as high as possible. In that sense it is perfectly logical to use criteria that are SMART for your label. SMART stands for Specific, Measurable, Assignable, Realistic, Time-related or some closely related variation on that description. There is a great article on Wikipedia on the topic: 'SMART Criteria', in *Wikipedia*, 15 May 2024, https://en.wikipedia.org/w/index.php?title=SMART_criteria&oldid=1223936953. . You can also find more info on the history of SMART goals at 'A Brief History of SMART Goals', Project Smart, 13 December 2014, https://www.projectsmart.co.uk/smart-goals/brief-history-of-smart-goals.php.

7  Rishi Bommasani et al., 'The Foundation Model Transparency Index v1.1 May 2024', May 2024.

| Company | Foundation Model | Index score (%) | Openness score label |
|---|---|---|---|
| Adept | Fuyu-8B | 33% | D |
| AI21 Labs | Jurassic-2 | 75% | B |
| Aleph Alpha | Luminous | 75% | B |
| Amazon | Titan Test Express | 41% | C |
| Anthropic | Claude 3 | 51% | C |
| BigCode/HF/ServiceNow | Starcoder | 85% | A |
| Google | Gemini 1.0 Ultra | 47% | C |
| IBM | Granite | 64% | B |
| Meta | Llama 2 | 60% | B |
| Microsoft | Phi-2 | 62% | B |
| Mistral | Mistral 7B | 55% | C |
| OpenAI | GPT-4 | 49% | C |
| Stability | Stable Video Diffusion | 58% | C |
| Writer | Palmyra-X | 56% | C |

Of course, foundational models are but one example of different tools that can be used in the context of generative AI use in a creative design context. However, since these foundational models are at the source of many other tools, they remain still quite relevant. So, if you want to pay attention to the ethical standing of the tools you use for your creative processes, it is something that can be taken into account without to much hassle. The index as he is constructed now, even has several elements that could be used in other labeling systems. For example the sustainability label we propose in the following paragraph, might use the openness score as part of its score determining mechanism.

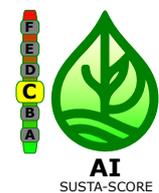

ChatGPT

We are still investigating what index could be used for a sustainability label, but are confident that using standard sustainability calculations (such as ESG, …) could be at the basis of that kind of label. The translation principle could be comparable to the one used for transparancy. A possible design is displayed here.

Labeling and reporting is a two-way street.

## AI ethics theories as lenses (think)

### Six lenses

There are several major ethical theories that can inform our approach to AI ethics. The theories mentioned here are certainly not the only possibilities. There is also often an overlap between these theories, certainly when one digs a bit deeper in the specific writings of the authors mentioned. The authors used in this article to illustrate the theory, are the main founders or early practitioners of it on the one hand, and some contemporary thinkers that try to fit the theory in a digitally enabled world on the other. The main reason to make the distinction between these theories, is the practicality of using them as lenses to look at the world in an ethically informed way.

Virtue ethics, traditionally associated with thinkers like Aristotle or Confucius, focuses on the moral character of individuals and the cultivation of virtues. According to Shannon Valor, one of the contemporary proponents of this theory, AI should enhance human virtues such as wisdom, care, and creativity rather than undermine them. This approach advocates for "technomoral futures" where technology development is value-driven [8].

Deontological ethics, rooted in the works of Immanuel Kant, emphasizes adherence to universal rules and the importance of duty. Ava Thomas Wright's concept of "rightful machines" underscores the importance of designing AI systems that operate within a framework of legal and moral obligations, ensuring their actions do not infringe upon others' freedoms [9].

---

8   Shannon Vallor, *Technology and the Virtues: A Philosophical Guide to a Future Worth Wanting* (New York: Oxford University Press, 2016).

9   Ava Thomas Wright, '8 Rightful Machines', in *Kant and Artificial Intelligence*, ed. Hyeongjoo Kim and Dieter

Consequentialist ethics, primarily known in its utilitarian form, with thinkers like Jeremy Bentham or John Stuart Mill, evaluates the moral worth of actions based on their outcomes. A current philosopher like Nick Bostrom highlights the importance of considering the long-term consequences of AI, advocating for the development of friendly AI to prevent catastrophic risks and ensure beneficial outcomes for humanity [10].

Contract ethics, originally linked to Hobbes, but in the modern age inspired by John Rawls and Robert Nozick [11], involves establishing ethical principles through a hypothetical social contract. Mathias Risse is a contemporary theorist who extends this framework to technology, emphasizing the need for fairness and justice in AI development, ensuring that AI technologies benefit all societal members equitably. He goes somewhat further than the typical contract ethics vision and tries incorporating Marxist and feminist ethics [12].

Care ethics, often connected to feminist theory, prioritizes human dignity and relationships. It's a rather recent development in ethical theory associated with people like Carol Gilligan and Nel Noddings [13]. In a certain way one could also consider Emmanuel Levinas someone close to the core principals of care ethics, although he is certainly not reducible to this school[14]. Virginia Dignum tries to apply the principals of care ethics in the context of AI systems. Respect for human dignity, with policies enforcing ethical standards is what she tries to realize in theory as well as in practice. This perspective aligns closely with deontological principles, emphasizing the role of policy in ensuring ethical AI [15].

Finally, existential ethics, rooted in the theories of Kierkegaard and Sartre, as proposed by Petros Terzis, emphasizes individual freedom and responsibility. In the context of AI, this approach suggests that individuals and organizations must navigate their interactions with AI systems responsibly, acknowledging the impact of their choices on their identity and society [16].

Using these six views to look at a practical design question, challenge or assignment, reinforces elements in the design thinking process (such as empathizing and problem definition formation) and opens new perspectives that not only make the solution more ethical, but also richer and more diverse. Integrating this in the design process doesn't make ethics an afterthought burdening creativity and innovation , but makes it part of the normal design process toolkit.

## *Zooming in and out*

Inspired by the double diamond design model, our proposed framework involves a process of zooming in and out on the ethical dimensions of AI in creative production processes. This model was inspired by the article of Luke Munn on "The uselessness of AI ethics" [17]. This somewhat provocative article doesn't explicitly refer to the double diamond or the creative process, but tries to show that using AI ethics

---

principles to make AI development more ethical is not only useless but even a dangerous distraction. One way to take on the problem of ethics in AI is described in an interview later as zooming in and zooming out[18]. Here we reuse this terminology in a similar way in the context of the creative process that makes use of AI.

Zooming out means looking at the bigger picture related to AI tools and their use. It tries to zoom out to a level where societal and environmental needs come in the picture. It involves considering the broader context and values at stake, drawing on the six ethical theories outlined above. In the examples in Table 1 this concerns the macro level. Although we are focusing here on the aspects related to the use of AI, this is often not only relevant in that context. The assignment, challenge or campaign itself, the diversity of the team, the governance structure of the company and other more general aspects surrounding the teams' work are part of this zooming out phase. Zooming out is looking at what you are doing with a helicopter view. The zooming out phase typically happens at the big transition moments in the double diamond process: at the start of the process, between the problem and the solution phase (at every iteration) and at the end of the process before concluding or deciding to re-iterate.

On the other hand, zooming in can occur more frequently and is looking at a more detailed level. In the table this is primarily described in the micro and meso level. It is mostly focusing on the effects of the tool use on the team itself and its direct stakeholders. In principle it could be done before and after each phase where AI tools are being used. It also very useful in the planning phase, when deciding which tools to use for the task. It can also be called upon by every participant in the creative process, team members as well as external stakeholders. In case there is a conductor for the team (as described in the The Team section), it is his responsibility to detect the necessity for a zooming in stop. Every zooming in moment can mean conducting small experiments to test the effects of AI interventions, gathering feedback from users, and iterating based on the results, but it could also involve doing some extra research, it all depends on the concrete case.

The table below gives an overview of some typical questions that might come up in a creative process using AI tools or integrating AI tools in a solution.

| Ethical Theory | Level | Creative Design Agency | Industrial Product Design Company | Marketing Agency |
|---|---|---|---|---|
| Virtue Ethics | Micro | How does the use of AI in our design tools foster creativity and empathy within the design team? | How does AI integration in our products encourage responsible usage among our designers? | How does the use of AI in marketing foster virtues like honesty and transparency within our team? |
| | Meso | Does integrating AI into our projects enhance virtues in our relationships with clients and suppliers? | Does AI in our products enhance the well-being of our customers and suppliers? | Are AI-driven campaigns enhancing virtues in our client relationships and partnerships? |
| | Macro | Are AI-driven design processes promoting ethical values in society? | Are we prioritizing virtues like reliability and user-friendliness in AI for societal benefit? | Are we ensuring our AI-driven messages foster societal respect and care? |
| Deontological Ethics | Micro | Are we ensuring our AI design tools adhere to ethical standards within the team? | Are our AI algorithms and data ethically used by our designers? | Are our AI-generated advertisements truthful within our team's practices? |
| | Meso | Does AI in our designs respect the rights of our clients and suppliers? | Does AI integration comply with safety standards affecting clients and suppliers? | Do our AI marketing practices respect privacy in client and partner relationships? |
| | Macro | Are our AI processes inclusive and accessible to all | Are we ensuring our AI-enhanced products do not harm society or | Are we ensuring our AI-driven marketing does not exploit |

---

18  *The Uselessness of AI Ethics*, 2022, https://www.youtube.com/watch?v=pfgYNxD0ndQ.

| | | | | |
|---|---|---|---|---|
| | | | societal groups? | the environment? | vulnerable societal groups? |
| Consequentialist Ethics | Micro | What are the potential positive and negative outcomes of using AI for our design team? | What are the potential consequences of AI on our design team's workflow? | What are the outcomes of using AI in marketing on our team's efficiency? |
| | Meso | How does AI maximize benefits for our clients and partners? | How can we minimize negative outcomes and maximize AI benefits for clients and suppliers? | How can AI in our campaigns maximize benefits and minimize harms for clients and partners? |
| | Macro | Are we considering long-term societal impacts of AI in our designs? | Are we considering AI's lifecycle impact on the environment? | Are we evaluating long-term societal effects of AI in marketing strategies? |
| Contract Ethics | Micro | Are we involving diverse team members in AI integration to ensure fairness? | Are we engaging our design team to ensure AI in our products meets ethical standards? | Are our AI marketing strategies fair within our team? |
| | Meso | Does AI in our designs meet the ethical standards of our clients and suppliers? | Does AI in product design address needs and rights of clients and suppliers? | Does AI in our campaigns respect the ethical standards of our clients? |
| | Macro | Are AI benefits equitably distributed across society? | Are we promoting fairness and justice in AI production for society? | Are we considering ethical implications of AI in marketing for society at large? |
| Care Ethics | Micro | Does AI in our design tools promote well-being and care within our team? | Are AI features in our products enhancing the quality of life for our design team? | Does AI in our marketing prioritize well-being within our team? |
| | Meso | Are we considering relational impacts of AI on clients and suppliers? | How do AI-driven practices reflect care and responsibility towards clients and suppliers? | How are we addressing relational impacts of AI on our clients in campaigns? |
| | Macro | How does our AI design promote empathy and compassion in society? | Are we ensuring AI in our products supports societal relationships? | Are we promoting messages that foster societal care and respect through AI? |
| Existential Ethics | Micro | Are we allowing freedom and individuality in AI-driven design tools for our team? | Does AI in our product design encourage responsible actions by our team? | Are AI campaigns promoting individual freedom within our team? |
| | Meso | How does AI empower clients and partners to make responsible choices? | How does AI design reflect freedom and autonomy of our clients and suppliers? | How are we encouraging responsibility in client choices with AI? |
| | Macro | Are we addressing the existential impact of AI on societal identity? | Are we considering existential implications of AI on societal lives? | Are we reflecting on the existential impact of AI marketing on societal identity? |

*Table 1: Zooming in and out*

It's important to realize that these questions are not a final fixed list of possible questions. In some circumstances some theories will be more relevant than in others. Within a theory, other questions or variations of questions can be formulated and they always have to be made concrete. These are just lenses and through these lenses one gets another view on the problem or assignment at hand. These views help guide the creative process in one direction or another and should never lead to a definitive

"yes/no" or "black/white" answer. They should make you pause and take some time to anticipate or reflect.

Anticipation involves proactively considering the potential consequences of AI systems before they are deployed, and taking steps to mitigate risks and harms. Reflection involves looking back on the impacts of AI systems after they have been implemented, and learning from successes and failures. By building anticipation and reflection into the design process, practitioners can cultivate a more ethically engaged and responsible approach to AI.

The big anticipation and reflection moments can be identified in the double diamond process on the seizures before and after each diamond.. The zooming out circles in the image below illustrate these moments. The smaller zooming in circles indicate moments of anticipation and reflection on a smaller scale and can happen at any time.

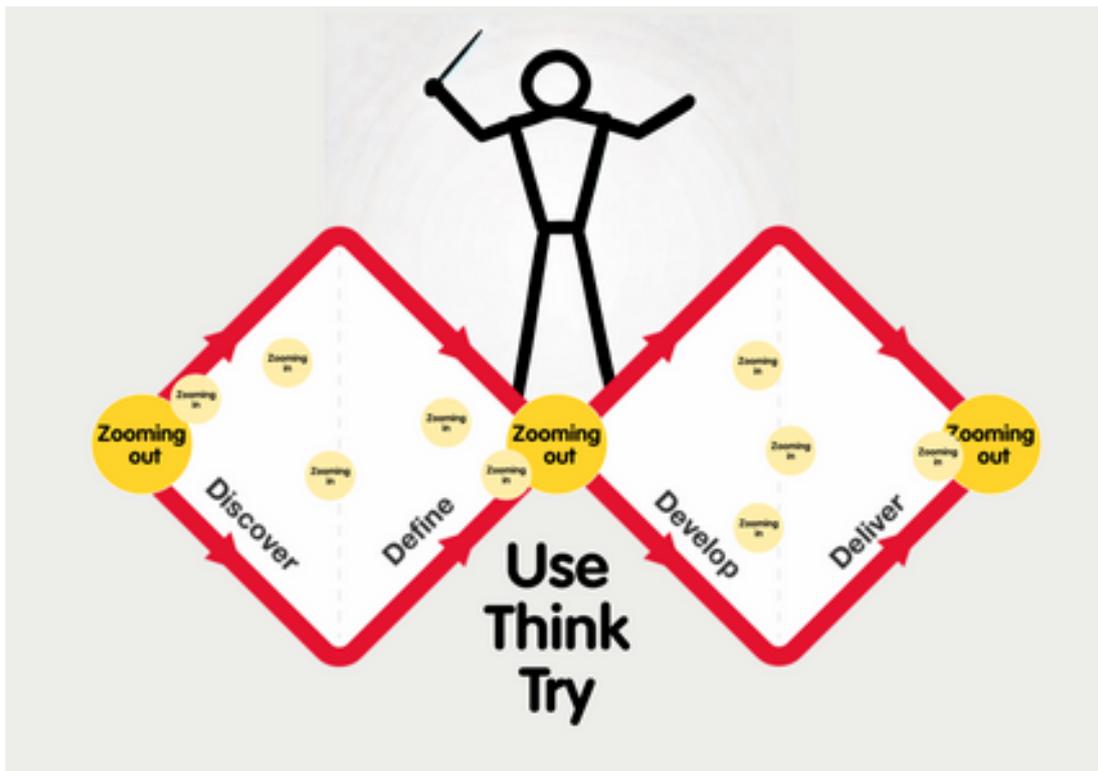

### *Contstraints as enablers*

Although it might seem that in the use and think phase we have only added constraints, this doesn't have to mean that everything becomes more difficult. Creativity and problem solving can thrive when boundaries emerge, given the right circumstances. It's by rubbing one constraint against another, that new possibilities and opportunities arise. For teams able to accept and embrace these constraints, there is freedom in them[19].

## Playful exploration (try)

Finally, our framework encourages a spirit of playful exploration in the use of AI for creative production. Rather than viewing AI only as a tool for optimizing efficiency or replacing human judgment, we suggest approaching it with a sense of curiosity, experimentation, and open-ended inquiry. This might involve using AI to generate unexpected ideas or challenge assumptions, while remaining grounded in fundamental ethical principles. It also means, using AI to do small scale experiments before relying on it

---

for real production efforts.

By adopting a playful mindset, the many ethically questionable aspects of AI can be used to open up our view of the world. Knowing that images, texts and concepts produced by generative AI are infested with bias, unfairness, opacity, misinformation and hallucinations can be used as a kind of mirror of ourselves. The recent book "The AI Mirror: How to Reclaim Our Humanity in an Age of Machine Thinking" by Shannon Vallor seems to suggest this too[20]. The current generation of generative AI is not the onset of beginning general intelligence. It is however a multi-faceted mirror reflecting our own good and bad traits.

Using AI realizing this conundrum opens possibilities of creative exploration that previously weren't there. Just as Tyler Reigeluth describes[21], looking at generative AI as just a way to promote efficiency and forgetting that it's in first instance a wonderful playing machine, leads us in a direction we definitely do not want. Machines and AI are not here to replace us, but need to be used as playmates (without the erotic innuendo). In that guise they are creativity enhancing tools that, when used properly, can help deliver ethically sound solutions to ethically sound tasks and problems.

## Conclusion

As AI technologies become increasingly integrated into creative production processes, it is essential for practitioners to develop a framework for navigating the ethical challenges and opportunities they present. By drawing on diverse ethical theories and emphasizing the importance of anticipation, reflection, and playful exploration, our proposed framework offers a practical approach for using AI in a responsible and value-aligned way. While there are no easy answers to the complex ethical questions raised by AI, we believe that by engaging with these issues in a proactive and principled manner, the design community can play a leading role in shaping the future of these powerful technologies.

---

20  Shannon Vallor, *The AI Mirror: How to Reclaim Our Humanity in an Age of Machine Thinking* (Oxford University Press, 2024).
21  Tyler Reigeluth, 'Play It Again: A Genealogy for Machine Learning', *Journal of Human-Technology Relations* 1 (12 June 2023), https://doi.org/10.59490/jhtr.2023.1.7015.